\documentclass[twocolumn,aps,prd,superscriptaddress,preprintnumbers,nofootinbib,10pt,floatfix]{revtex4-2}

\usepackage{amssymb}
\usepackage{amsmath}
\usepackage{slashed}
\usepackage[colorlinks]{hyperref}
\usepackage{tablefootnote}
\usepackage{verbatim}
\usepackage{graphicx}

\newcommand{\Sp}[1]{\ensuremath{\text{Sp}(#1)}}
\newcommand{\SU}[1]{\ensuremath{\text{SU}(#1)}}
\newcommand{\SO}[1]{\ensuremath{\text{SO}(#1)}}
\newcommand{\Nfcrit}{\ensuremath{N_f^\text{crit}}}

\begin{document}

\title{On the dependence of the Landau gauge ghost-gluon-vertex on the number of flavors}

\author{Fabian Zierler}
\email{fabian.zierler@swansea.ac.uk}
\affiliation{Department of Physics, Faculty of Science and Engineering, Swansea University, Singleton Park, SA2 8PP Swansea, Wales, United Kingdom}
\affiliation{Institute of Physics, University of Graz, NAWI Graz, Universit\"atsplatz 5, 8010 Graz, Austria}
\author{Reinhard Alkofer}
\email{reinhard.alkofer@uni-graz.at}
\affiliation{Institute of Physics, University of Graz, NAWI Graz, Universit\"atsplatz 5, 8010 Graz, Austria}

\begin{abstract}
The gauge-boson, ghost and fermion propagators as well as the gauge-boson--ghost vertex function are studied for \SU{N}, \Sp{2N} and \SO{N} gauge groups. We solve a set of coupled Dyson-Schwinger equations in Landau gauge for a variable, fractional number $N_f$ of massless fermions in the fundamental representation. For large $N_f$ we find a phase transition from a chirally broken into a chirally symmetric phase that is consistent with the behavior expected inside the conformal window. Even in the presence of fermions the gauge-boson--ghost-vertex dressing remains small. In the conformal window this vertex shows the expected power law behavior. It does not assume its tree-level value in the far infrared, but the respective dressing function is a constant greater than one.
\end{abstract}

\maketitle

\section{Introduction}
\label{sec:intro}
Strongly interacting gauge theories with fermions are both a cornerstone in the Standard Model (SM) of particle physics in the form of Quantum Chromodynamics (QCD) and provide a multitude of Beyond the Standard Model (BSM) models. Due to asymptotic freedom \cite{Gross:1973id,Politzer:1973fx} they are automatically ultraviolet (UV) complete and can be treated perturbatively at high energies. In addition, the perturbative two-loop beta function of confining asymptotically free gauge theories such as QCD suggests the existence of an infrared (IR) conformal fixed-point \cite{Banks:1981nn, Caswell:1974gg} for a sufficient number of massless fermions \Nfcrit. These (near-)conformal theories have gained some interest as candidate theories for a walking technicolor scenario, {see, {\it e.g.}, \cite{Cacciapaglia:2020kgq,Aarts:2023vsf} for recent reviews.} Apart from that confining gauge theories with several fermions are also considered in the context of Dark Matter {\cite{Cline:2013zca,Hochberg:2014kqa}} models in different gauge groups and even using different fermion representations. Due to the confining nature of such theories a study of the mid- to low-energy regime of strongly interacting theories as well as bound-state properties necessitates the use of non-perturbative methods such as lattice field theory or functional methods like the functional renormalization group (FRG) or Dyson-Schwinger equations (DSEs). Lattice results on the onset of the conformal window are mostly available for \SU{2} and \SU{3} gauge theories with fermions in the fundamental and adjoint representation (see, {\it e.g.}, \cite{DeGrand:2015zxa, Witzel:2019jbe} for reviews) and the exact values of \Nfcrit\ are often still debated. Due to the high computational cost for light fermions as well large color groups and fermion representations an exhaustive study of different gauge groups and fermion representations is (currently) prohibitively expensive using lattice methods. Additionally, and maybe even more importantly, theories with a walking coupling constant require to consider quantities at vastly different scales, {\it cf.} the discussion in the review \cite{DeGrand:2015zxa}.

Functional methods have so far explored \SU{3} gauge theories. Within DSEs a study of the Landau gauge propagators in \SU{3} has been performed \cite{Hopfer:2014zna}. A value of $\Nfcrit \approx 4.5$ has been obtained which is at odds with other methods. This has been attributed to lack of a self-consistent inclusion of vertex functions,
in particular of the quark-gluon vertex. 
Using the FRG a value of $\Nfcrit = 10.0^{+1.84}_{-0.92}$  \cite{Gies:2005as} has been obtained. This is consistent with many lattice studies as well as a perturbative determination using the conformal expansion in $\Delta_{N_f} = N_f^\text{AF} - N_f$ \cite{Lee:2020ihn} up to fourth order {and purely analytical methods \cite{Dietrich:2006cm, Sannino:2009aw, Armoni:2009jn}}. Hereby $N_f^\text{AF}$ is the number of fermion flavors at which asymptotic freedom is lost. These methods have also been applied to \SU{N}, \Sp{2N} and \SO{N} gauge groups with fermions in different representations.

Spectral functions of propagators have been studied in gauge theories with a Banks-Zaks fixed point using perturbation theory 
up to five-loop order and Callan-Symanzik resummations \cite{Kluth:2022wgh}. Hereby a 
significant dependence on the value of the gauge coupling and 
on the employed order of perturbation theory has 
been found.
The scheme and gauge dependence of the existence of Banks-Zaks fixed points for a SU(3) gauge theory and fundamentally charged fermions 
has been studied up to five loop order in \cite{Gracey:2023unc}
with the result that the existence of Banks-Zaks fixed points
is quite stable. The studies \cite{Kluth:2022wgh,Gracey:2023unc} provide impressive progress on 
the existence of the conformal window and of properties of correlation
functions in the upper region of the conformal window.

In principle, a reliable determination of 
correlation functions from DSEs, and subsequently of the spectrum by employing bound-state equations, close to and in the lower region
of the conformal window
is highly desirable for BSM physics: A system of truncated DSEs is systematically improvable by including higher order correlation functions, {\it cf.} ref.\ \cite{Pawlowski:2022oyq}, and can be solved using modest computational resources. It allows us to determine the correlation functions (the complete set of which encodes the entire physics of any given theory) over a large range of momenta. The number of colors $N_c$ and fermion flavors $N_f$ enters in the form of Casimir invariants which allows to easily change the color group and fermion number. Additionally, we can directly study the system of equations even for non-integer $N_c$ and $N_f$. 

For \SU{3} Yang-Mills theory a model-parameter free system of all primitively divergent correlation functions was solved in \cite{Huber:2020keu, Huber:2020ngt}. Glueball masses have been predicted, and this in good quantitative agreement with lattice results where available \cite{Athenodorou:2020ani}. 
However, no such calculation has been performed which includes the fermion sector at this level of sophistication. Another promising approach is to use lattice input in order to solve the DSEs self-consistently \cite{Gao:2021wun,Aguilar:2021okw}. These calculations, however, rely on the availability of dedicated, computationally expensive lattice results. Nevertheless, in particular for the ghost sector of quenched QCD this 
approach has proven the consistency of the different computational schemes with each other and with the physical concept of ghost dominance in quenched Landau gauge QCD \cite{Aguilar:2021okw}.

We therefore set out to study a system consisting of the Landau gauge quark, gluon and ghost propagator as well as the ghost-gluon vertex\footnote{Without loss of generality we will use from here on a naming scheme borrowed from QCD.}. We solve a truncated set of DSEs simultaneously using models for the three-gluon-vertex and the quark-gluon-vertex for different gauge groups. This has multiple advantages compared to the investigation reported in Ref.~\cite{Hopfer:2014zna}: Its shortcomings were attributed to the exclusion of self-consistent vertex function, most notably the quark-gluon vertex. By including the ghost-gluon vertex self-consistently instead of taking it to be bare we can for the first time test if a vertex function shows the expected IR behavior in the conformal window. This study does not only serve the purpose of investigating the influence of the ghost-gluon vertex but, also is a first step towards a self-consistent calculation of other vertex functions, most notably the quark-gluon vertex and the three-gluon vertex close and in the conformal window. 

The Landau gauge ghost-gluon-vertex has been studied on the lattice in $SU(3)$ gauge theory outside the conformal window \cite{Cucchieri:2008qm, Ilgenfritz:2006he}. The qualitative behavior is consistent with the one seen in Dyson-Schwinger studies.  Additionally, the vertex was studied in $SU(2)$ with two adjoint fermions \cite{Maas:2011jf} in the 'minimal walking technicolor' scenario. Within uncertainties, it did not show any momentum dependence and is consistent with a tree-level dressing. The propagators of $SU(2)$ gauge theory with $N_f=1/2,1,3/2$ and $2$ fermions (corresponding to an integer number of Weyl fermions) were studied on the lattice in \cite{Bergner:2017ytp}. The qualitative behavior of the propagators inside the conformal window was found to be consistent with the behavior seen in the Dyson-Schwinger study presented in ref.\  \cite{Hopfer:2014zna}.

Employing functional methods it has been shown in \cite{Schleifenbaum:2004id, Huber:2012kd} that the bare approximation for the ghost-gluon vertex is well motivated in the Yang-Mills theory. A similar investigation for non-vanishing $N_f$ has not yet been performed. Additionally, the simultaneous investigation of multiple gauge groups will enable us to compare the different gauge theories relative to another. This can also be seen as an extension of the quark DSE studied for larger gauge groups in \cite{Llanes-Estrada:2018azk} to a truncation involving propagators of the associated Yang-Mills sector and variable fermionic content.
 
\section{System of Dyson-Schwinger equations}
\label{sec:dses}

All expressions in the following are understood to be formulated in
Euclidean momentum space, {\it i.e.}, after a Wick rotation.

\begin{figure}
    \begin{align*}
    \includegraphics[scale=0.18]{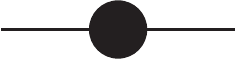}^{-1} = & ~ 
    \includegraphics[scale=0.18]{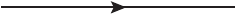}^{-1} +   
    C_F ~ \hbox{\includegraphics[scale=0.18]{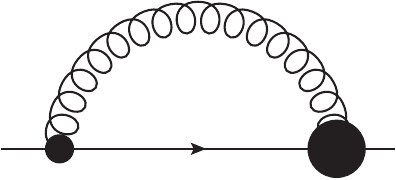}} & \\[5pt]
    \includegraphics[scale=0.18]{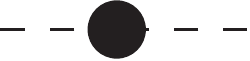}^{-1} = & ~ 
    \includegraphics[scale=0.18]{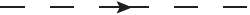}^{-1} +   
    C_A ~ \hbox{\includegraphics[scale=0.18]{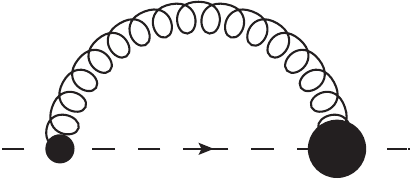}} & \\[5pt]
    \vcenter{\hbox{\includegraphics[scale=0.18]{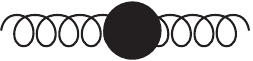}}}^{-1} = & ~ 
    \vcenter{\hbox{\includegraphics[scale=0.18]{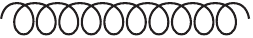}}}^{-1} +  
    {T N_f} ~ \vcenter{\hbox{\includegraphics[scale=0.18]{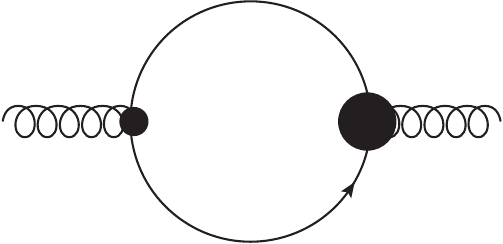}}} & \\ 
    & +  C_A \left( \vcenter{\hbox{\includegraphics[scale=0.18]{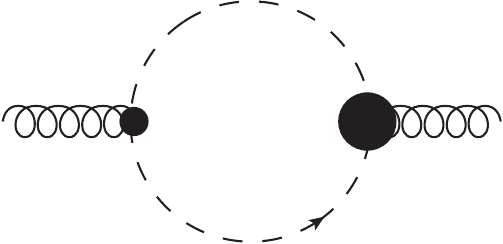}}} +  
    \vcenter{\hbox{\includegraphics[scale=0.18]{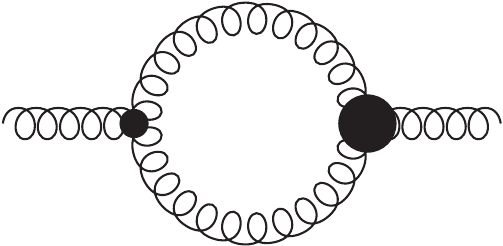}}} \right) \\[5pt]
    \vcenter{\hbox{\includegraphics[scale=0.18]{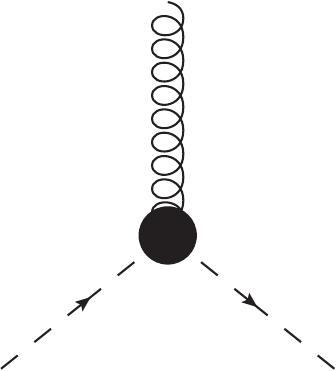}}} = & ~
    \vcenter{\hbox{\includegraphics[scale=0.18]{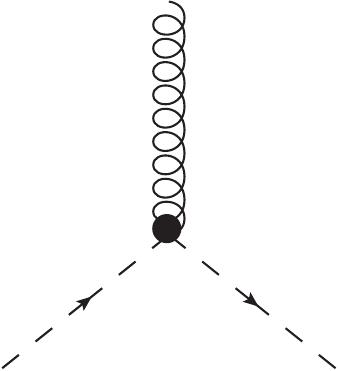}}}     
    + {\frac 1 2 C_A^2} 
    \left(\vcenter{\hbox{\includegraphics[scale=0.18]{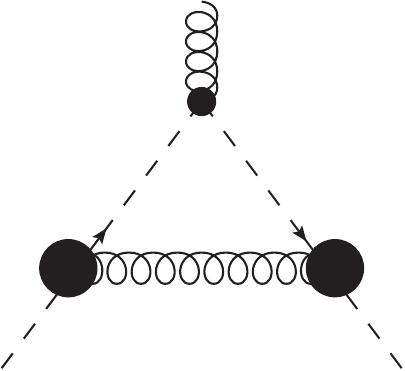}}} \!+\!
    \vcenter{\hbox{\includegraphics[scale=0.18]{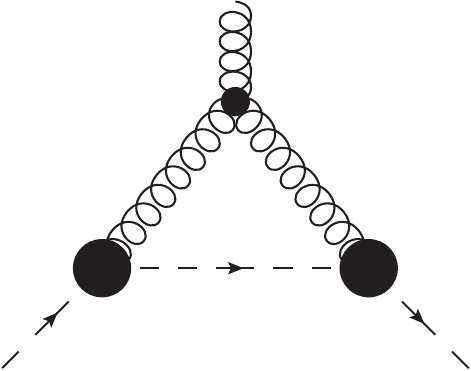}}} \right) &
    \end{align*}
    \caption{Truncated DSEs for the propagators (quark, ghost and gluon from top to bottom) and the ghost-gluon vertex. Thick blobs denote dressed vertices whereas small blobs denote bare vertices. All internal lines are fully dressed. The $N_c$-dependent prefactors for quarks in the fundamental
    representation are given in table \ref{tab:Casimirs}.}
    \label{fig:GGV_DSE}
\end{figure}

In figure \ref{fig:GGV_DSE} we depict the truncated system of DSEs considered in the following for an arbitrary gauge group and arbitrary fermion representation. For \SU{3} gauge theory, the same system of propagators without the inclusion of the ghost-gluon-vertex has been studied in the past \cite{Fischer:2003rp,Fischer:2005en,Hopfer:2014zna} (see also \cite{Cyrol:2017ewj} for an FRG calculation of both propagators and vertices involving quarks).
The color group and fermion representation only enter in the form of the contracted color matrices. The overall form of the DSEs is always the same since the corresponding Lagrangian is always
\begin{align}
    \label{eq:Lagrangian}
    \mathcal L &= \frac{1}{2} \text{Tr}\left[ F_{\mu\nu} F^{\mu\nu}\right] + \sum_\text{flavors} \bar q_f \left( - \slashed D + m \right) q_f 
\end{align}
with the covariant derivative $D_\mu = \partial_\mu + i g  A_\mu$ and the field strength tensor $F_{\mu\nu} = \partial_\mu A_\nu - \partial_\nu A_\mu + ig \left[ A_\mu, A_\nu \right]$. The fermion fields of different flavors are denoted by $q_f$. As we are interested in the infrared fixed point we will always set the bare quark mass to zero, $m=0$. For any two theories only the gauge group and thus representation of the gauge fields $A_\mu$  as well as the representation of the quark fields $q$ change. Herein, we will only consider quarks in the fundamental representation. The truncation of the ghost-gluon-vertex DSE is chosen such that no explicit model dependence enters the DSE, see \cite{Huber:2012kd}. We collect the different color prefactors appearing in the propagator DSEs in table \ref{tab:Casimirs}. 
\begin{table}
    \centering
    \begin{tabular}{|c|c|c|c|}
      \hline 
              & $T$   & $C_F$          & $ C_A $ \\ \hline \hline
      \SU{N}  & $1/2$ & $(N^2-1)/(2N)$ & $N$ \\ \hline  
      \Sp{2N} & $1/2$ & $(2N+1)/4 $    & $N+1$ \\ \hline  
      \SO{N}  & $1$   & $(N-1)/2 $     & $N-2$ \\ \hline \hline 
    \end{tabular}
    \caption{The Casimir invariants $C_F$ and $C_A$ as well as the generator normalization $T$ 
    which determine the prefactors of the individual diagrams in the coupled set of propagator DSEs,
    {\it cf.} figure~\ref{fig:GGV_DSE}. }
    \label{tab:Casimirs}
\end{table}
The Landau gauge gluon $D_{\mu\nu}(p)$ and ghost propagator $D_G(p)$ are fully parameterized by one scalar dressing function whereas the quark propagator $S(p)$ and the ghost-gluon vertex 
$\Gamma^{A\bar c c}_\mu(k=-p-q;p,q)$ are described by two dressing functions each
\begin{align}
    &D_{\mu\nu}(p) =  \left( \delta_{\mu \nu} - \frac{p_\mu p_\nu}{p^2} \right) \frac{Z(p^2)}{p^2}\, ,\\
    &D_G(p) = - \frac{G(p^2)}{p^2} \, ,\\
    &S_f(p) = i \slashed p A_f(p^2)+ B_f(p^2) \, ,\\
    &\Gamma^{A\bar c c}_{\mu,abc}(k;p,q) = igf^{abc} \times \nonumber\\
    &~\hspace{20mm} \times  \left(A(k;p,q) p_\mu + B(k;p,q) k_\mu \right) \, .
\end{align}
In the ghost-gluon vertex we denote the gluon momentum by $k$, the ghost momentum by $q$ and the anti-ghost momentum by $p$. We can parameterize the vertex dressing functions by the squared momenta $k^2$ and $p^2$ as well as the angle $\theta$ between the gluon and anti-ghost momentum. The ghost momentum squared is then $q^2 = p^2 + k^2 + 2 \sqrt{p^2 \cdot k^2} \cos(\theta)$. In Landau gauge the DSE for the ghost-gluon-vertex' dressing function $A(k^2,p^2,\cos(\theta))$ does not depend on $B(k^2,p^2,\cos(\theta))$ which is vanishing at tree level and does not enter the propagator DSEs. We therefore only calculate $A(k^2,p^2,\cos(\theta))$, {\it i.e.}, the coefficient of $p_\mu$, as this is sufficient to check whether its qualitative behavior changes across the phase transition. For later use we define, as usual, the quark mass function as $M_f(p^2) = B_f(p^2)/A_f(p^2)$. For the three-gluon- and quark-gluon-vertex we employ the models of \cite{Hopfer:2014zna} that consist of the tree-level tensor structures times a dressing function.

\section{Numerical scheme} \label{sec:numerics}
We numerically solve the system by an iterative procedure. We use a global Newton-Raphson method \cite{Maas:2005xh} for the Yang-Mills propagators and a local fixed point iteration for the quark dressing functions. If the ghost-gluon-vertex is taken to be dynamical, we perform a single iteration step for the vertex between every update of the propagators -- otherwise the system will not converge as was noted in \cite{Schleifenbaum:2004id}. We then repeat this macrocycle until the overall system of propagators and vertex converges. Using relaxation when updating the quark dressing functions for the Yang-Mills iteration lead to better convergence of the macrocycle when changing $N_f$. We find that the numerical parameters reported in \cite{Huber:2012kd} for the ghost-gluon vertex in pure Yang-Mills theory are sufficient for $N_f>0$.  We follow \cite{Hopfer:2014zna} in their choice of renormalization, scale setting and the subtraction of spurious quadratic divergences in the gluon DSE. Hereby the importance of a sufficiently good radial integration for the propagators close to the phase transition is noteworthy. We found that depending on the gauge group up to $4~000$ integration nodes using a logarithmically spaced Gauss-Legendre quadrature were needed close to the chirally symmetric phase.

A few words on our choice for scale setting are in order. For \SU{N} and \Sp{2N} gauge theory, we determine the value of the coupling $g^2$ at the renormalization point for $N_f \leq 2$ by fixing the value of the pion decay constant using the approximation \cite{Pagels:1979hd, Hopfer:2014zna}
\begin{align}
    f_\pi^2 = \frac{Z_2 \dim(R)}{\pi^2} \int \!\! dq^2 \frac{q^2 M_f(q^2) \left(M_f(q^2)+\frac{q^2}{2}M_f'(q^2)\right) }{(q^2+M_f^2(q^2))^2A_f(q^2)}  \, ,   
\end{align}
and then extrapolate up to quadratic order in $N_f$ for $N_f > 2$. 
Hereby, $Z_2$ is the quark wave function renormalization constant, and $\dim(R)$ is the dimensionality of the quark representation, i.e. $N$ for \SU{N} and \SO{N} gauge theories and $2N$ for a \Sp{2N} gauge theory.

In \cite{Hopfer:2014zna} it was found that the particular choice of the scale setting procedure does not influence $N^{\rm crit}_f$ but can lead to small quantitative differences below the phase transition. For \SO{N} we found it numerically advantageous to fix the value of the quark mass function $M_f$ in the infrared limit to a value of $M_f(p^2 \to 0)=0.5$ {\it a.u.}\footnote{The units are in this case, of course, arbitrary and purely internal. This specific value has been chosen to be close to the QCD value in GeV.} For any given theory, we determine $g^2$ from the set of propagator DSEs and leave the bare coupling unchanged once we introduce the dynamical ghost-gluon-vertex. This allows us to study the effects of the vertex alone.

For chirally broken phases a family solution is known, which is parameterized by the value $G(0)$. It either assumes a finite, constant value and the associated solutions are known as \textit{decoupling} type solutions, or it diverges and the Yang-Mills propagators obey a power-law behavior which is typically referred to as \textit{scaling} solution. In the conformal window only the scaling solution is expected to exist, due to the absence of any scales in an IR conformal theory. This is particularly relevant for quark-gluon-vertex in our truncation, which is modelled by
\begin{align}
    \label{eq:quark-gluon-vertex-model}
    \Gamma^{A\bar q q}_{\mu,abc}(k;p,q) &= \Gamma^{(0),A\bar q q}_{\mu,abc}(k;p,q) D^{A \bar q q}(k;p,q) \\
    D^{A \bar q q}(k;p,q) &= \tilde Z_3 G(k^2)^2 ~ \frac{A(p^2) + A(q^2)}{2} \nonumber, 
\end{align}
with $\tilde Z_3$ being the ghost wave function renormalization constant.
This quark-gluon vertex model thus enhances the interaction strength between quarks and gluons depending on the value of $G(k^2 \to 0)$.
In light of this, we identify the onset of the conformal window by the $N_f$ for which the scaling type solution, which maximizes the infrared quark-gluon interaction, does not lead to any generation of a dynamical quark mass. In practice, we require that the quark mass function vanishes within machine precision such that $M_f(p^2) < \epsilon = 10^{-14}$ for all momenta $p$.

\section{Results}
\label{sec:results}

We first consider \SU{3}\ gauge theory with fundamental fermions, where results for the truncated system without the ghost-gluon vertex are already available. We discuss the qualitative behavior of the correlation functions  -- and the ghost-gluon vertex in particular -- in section \ref{sec:results_vertex}. The vertex is of even lesser importance than in Yang-Mills theory where the bare approximation is already deemed to be well justified. Then we look in section \ref{sec:loops_gluonDSE} at the interplay between the loops in the gluon DSE where the number of quark flavors $N_f$ enters explicitly. We discuss other $\SU{N}$, $\Sp{2N}$ and $\SO{N}$ gauge groups in section \ref{sec:other_groups}.

\subsection{QCD propagators and ghost-gluon vertex}
\label{sec:results_vertex}

\begin{figure}
    \centering
    \includegraphics[width=.41\textwidth]{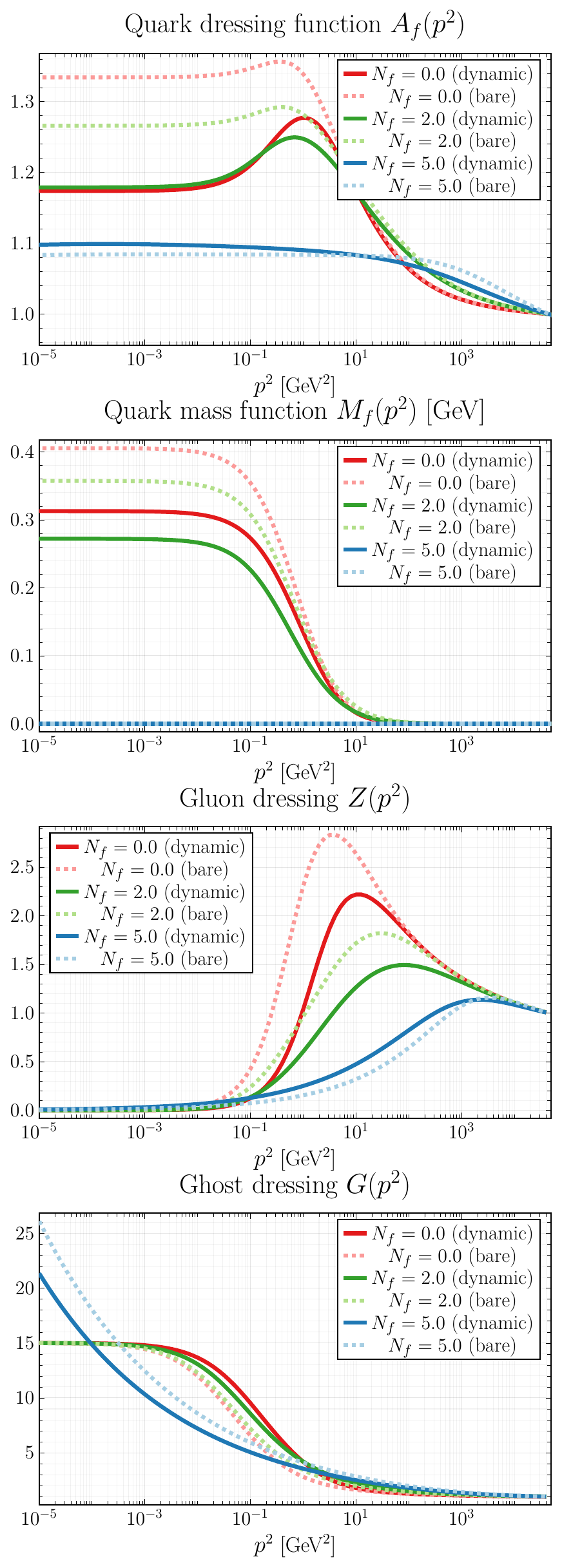}
    \caption{Dressing functions of the propagators for different $N_f$ with and without a dynamical ghost-gluon vertex in QCD obtained from the truncated set of DSEs. The quark mass function's value at vanishing momenta, $M_f(p^2 \to 0)$, is an order parameter for chiral symmetry breaking. For $N_f=5$ chiral symmetry is restored, and the Yang-Mills propagators obey power laws in the far IR. }
    \label{fig:prop_GGV_noGGV}
\end{figure}

\begin{figure}
    \centering
    \includegraphics[width=.47\textwidth]{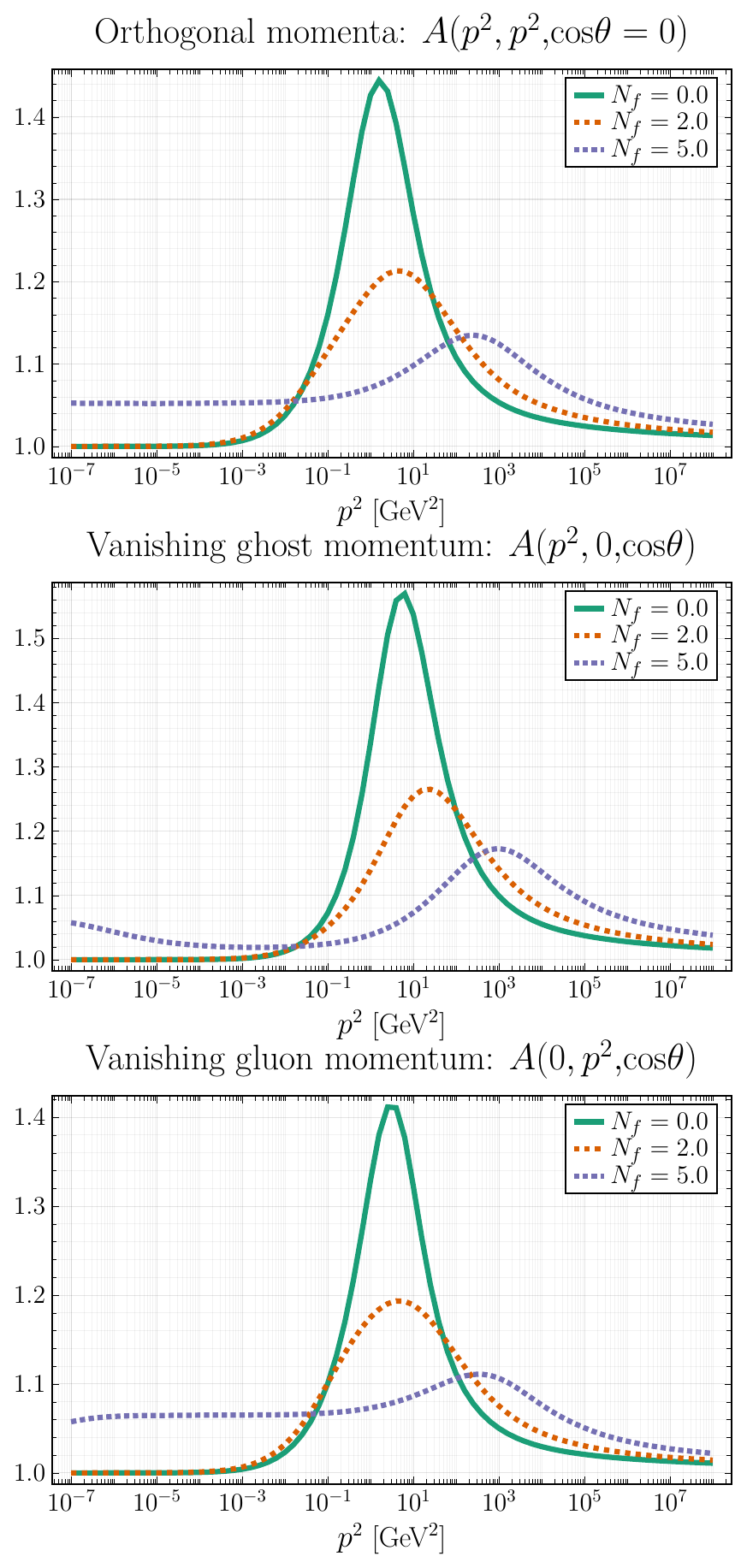}
    \caption{Ghost-gluon vertex for $N_f=0,2,5$ in QCD obtained from the truncated set of DSEs. The larger the number of quarks the closer the vertex' dressing function gets to its tree-level value of $1$. For $N_f=5$ the soft-limit value of the vertex does not assume its tree-level value one.}
    \label{fig:GGV_vs_Nf}
\end{figure}

As is clearly visible from the propagators' dressing function displayed in Fig.\ \ref{fig:prop_GGV_noGGV} the already small influence of ghost-gluon vertex at $N_f=0$ decreases as $N_f$ is increased. This is easily understood from the fact that the ghost-gluon vertex dressing function $A(k^2,p^2,\cos(\theta))$ moves closer to its tree-level value, see Fig.\ \ref{fig:GGV_vs_Nf}. Above the phase transition the vertex dressing function depends only mildly on the momenta. {This is consistent with behavior seen on the lattice in $\SU{2}$ gauge theory with two adjoint fermions \cite{Maas:2011jf} as well as $\SU{3}$ gauge theory with $N_f=12$ \cite{Maas:2023private}.} Note that the corresponding soft-limit value of the vertex does not assume its tree-level value one, see also the discussion below. 

{For \SU{3} we observe a critical value of $\Nfcrit=4.67$, with a numerical uncertainty of $\pm 0.02$, for both the system with a dynamical ghost-gluon-vertex and the system with a bare vertex. We do not observe any influence of the inclusion of the vertex on \Nfcrit. Our value deviates mildly from the results quoted in \cite{Hopfer:2014zna}. We attribute this to the more careful IR matching outlined in sec.~\ref{sec:numerics}}. 
Except this slight technical improvement the quantitative result of \Nfcrit\ is unchanged and still in significant tension with available results from lattice QFT, the FRG and perturbative calculations.  This indicates that our results can still only be seen as a study of the qualitative behavior of the correlation functions even with the ghost-gluon vertex included.

Above the phase transition the Yang-Mills dressing functions obey in the infrared (IR) power laws
\begin{align}
    \label{eq:IR_exponent}
    Z(p^2) &= a_Z \cdot (p^2)^{2\rho} \, , \nonumber \\ 
    G(p^2) &= a_G \cdot (p^2)^{-\rho} \, .
\end{align}
{The exponent $\rho$ does not significantly change after the inclusion of the ghost-gluon vertex. We were unable to resolve a difference within our numerical setup. With the purely numerical uncertainties given in parentheses, for the system of the propagators we determined a value of $\rho = 0.174(4)$ and a for the system with a dynamical vertex the exponent is $\rho = 0.178(4)$.} The propagators show the same qualitative behavior as in the case of the scaling type solution below the phase transition for which the critical exponent assumes a value $\approx 0.595$  \cite{Lerche:2002ep,Zwanziger:2001kw,Fischer:2002hna,Alkofer:2003jr}\footnote{{Note, that the exponents extracted from lattice gauge theory were found to be substantially smaller than the ones obtained from the DSE approach \cite{Ilgenfritz:2006he,Bergner:2017ytp}.}}. 
In the UV we recover the expected perturbative behavior given by
\begin{align}
    \label{eq:UV behavior}
    &Z(p^2) = Z(s^2) \left(\omega \log \left( \frac{p^2}{s^2}\right) + 1 \right)^\gamma,  \nonumber \\ 
    &G(p^2) = G(s^2) \left(\omega \log \left( \frac{p^2}{s^2}\right) + 1 \right)^\delta, \\ 
    &\omega = \left( \frac{11C_A - 4 T N_f}{3} \right)\left( \frac{g^2 G(s^2)^2Z(s^2)}{16 \pi^2} \right) \!,  \nonumber
\end{align}
and the anomalous dimension of the gluon $\gamma = (-13C_A + 8 T N_f)/(22C_A - 8 T N_f)$ and of the ghost $\delta = (-1-\gamma)/2$ \cite{Muta:2010xua}. 

With the given IR behavior for the propagators in the conformal window the ghost-gluon vertex function for symmetric momenta will in general assume a non-trivial constant in the IR \cite{Alkofer:2004it}. Indeed, we observe that the vertex' dressing function for vanishing symmetric momenta deviates from the tree-level value of $one$. This is in agreement  with the results of ref.\ \cite{Huber:2012kd} where the IR behavior for the ghost-gluon vertex has been studied in  Yang-Mills theory for the scaling type solution to DSEs. We conclude that the ghost-gluon vertex shows no signs of any unexpected behavior, and that using a bare approximation provides qualitatively correct results even for $N_f>0$. In addition, the behavior of the ghost-gluon vertex function in the far IR provides further numerical evidence for the existence of an IR fixed point above \Nfcrit. 

Although the quantitative result for \Nfcrit is at odds with the one obtained by other methods a further analysis of the quark loop contribution in the gluon DSE will shed some light on how the phase transition is reached, and results for different gauge groups will allow us to the compare the relative sizes of \Nfcrit , {\it{cf.}} sect.\ \ref{sec:other_groups}.

\begin{figure}
    \centering
    \includegraphics[width=.47\textwidth]{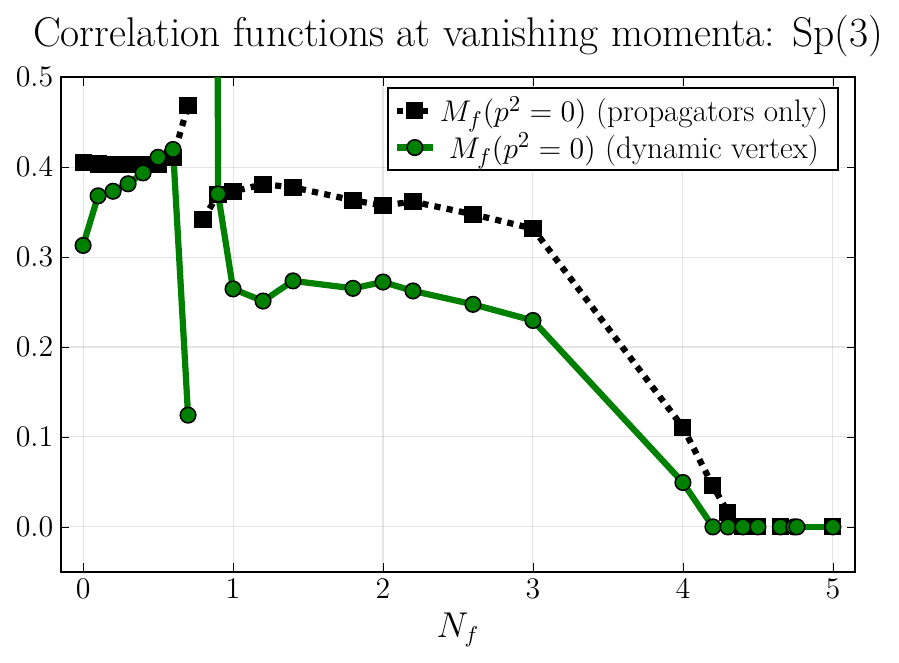}
    \caption{Infrared value of $M_f(p^2 \to 0)$ as a function of $N_f$. Below the onset of the conformal window we always show the decoupling type solutions which leads to a smaller value of $N_f$ for which no fermion masses are dynamically generated (see the discussion in Sec.~\ref{sec:numerics}). At low values of $N_f$ the system becomes numerically unstable. This can be attributed to cancellations of the loop diagrams in the gluon DSE -- see Sec.~\ref{sec:loops_gluonDSE}.}
    \label{fig:M0_vs_Nf}
\end{figure}

In Fig.~\ref{fig:M0_vs_Nf} we show $M_f(p^2 \to 0)$ as a function of $N_f$ with and without the dynamical ghost-gluon-vertex. At around $N_f \approx 0.8$ we see the effects off a cancellation of the loops in the gluon DSE as will be discussed in Sec.~\ref{sec:loops_gluonDSE}. For consistency, we always show values of the decoupling type solutions close to the conformal window. As noted in Sec.~\ref{sec:numerics} this will underestimate the transition into the conformal window. For the determination of $\Nfcrit$ we require that the scaling type solution do not lead to any effective mass generation. The overall value of $M_f(0)$ is reduced by the inclusion of a ghost-gluon vertex. This can be attributed to our scale setting procedure. We do not observe any qualitative changes that arise from the inclusion of the dynamical ghost-gluon-vertex.  

\subsection{\texorpdfstring{$N_f$}{Nf} dependence of the loop diagrams in the gluon DSE} 
\label{sec:loops_gluonDSE}

The only explicit appearance of the number of quarks $N_f$ is in the gluon DSE as the prefactor of the quark loop, see Fig.\ \ref{fig:GGV_DSE}, and only the gluon propagator and the quark-gluon-vertex enter the quark DSE. Therefore, it is reasonable to assume that the interplay between the three loops in the gluon DSE is at the heart of the phase transition. To this end compare the contributions of the different loops for converged solutions to the considered set of DSEs.

\begin{figure}
    \centering
    \includegraphics[width=.48\textwidth]{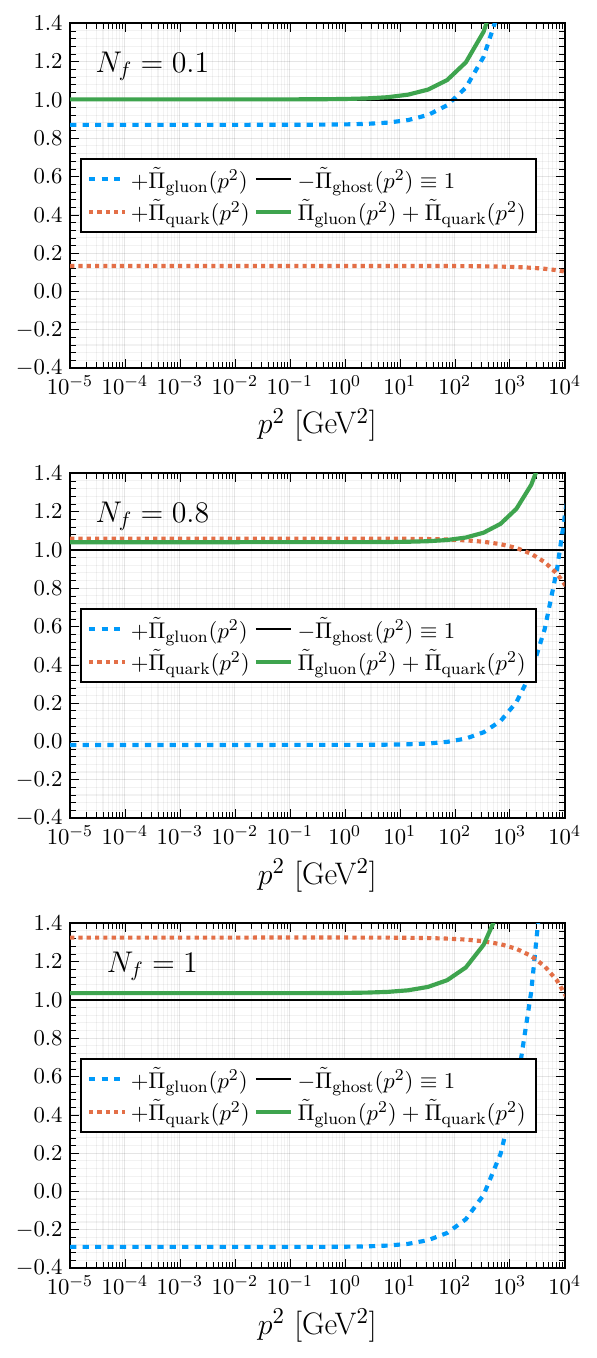}
    \caption{The gluon loop $\tilde{\Pi}_\text{gluon}(p^2) \equiv -\Pi_\text{gluon}(p^2)/\Pi_\text{ghost}(p^2)$ and quark loop $\tilde{\Pi}_\text{quark}(p^2) \equiv -\Pi_\text{quark}(p^2)/\Pi_\text{ghost}(p^2)$ of the gluon DSE, relative to the ghost loop, $\Pi_\text{ghost}(p^2)$ in $SU(3)$ with fundamental fermions. The dashed lines at one indicates the ghost loop. The sum of quark and gluon loop has large cancellations with the ghost loop. Even for small $N_f$ the quark loop becomes quickly significant. At $N_f \approx 0.8$ the gluon loop changes its sign in the infrared and mid-momentum regime.}
    \label{fig:loop_cancellation_new}
\end{figure}

The gluon propagator DSE, when written as an equation for the renormalization function $Z(p^2)$, possesses the following decomposition in quark, gluon and ghost loops:
\begin{align}
    Z(p^2)^{-1} = Z_3 + \Pi_\text{quark}(p^2) + \Pi_\text{gluon}(p^2) + \Pi_\text{ghost}(p^2).
\end{align}
For $N_f=0$, the gluon and the ghost loop vary over many orders of magnitude, they are  opposite in sign, and for small momenta they are of almost identical magnitude, in agreement with the behavior seen in \cite{Eichmann:2021zuv}. At momenta above several GeV the gluon loop dominates. In the IR region large cancellations occur, and thus we plot in Fig.\ \ref{fig:loop_cancellation_new} the ratios $\tilde{\Pi}_\text{gluon}(p^2) \equiv -\Pi_\text{gluon}(p^2)/\Pi_\text{ghost}(p^2)$ and $\tilde{\Pi}_\text{quark}(p^2) \equiv -\Pi_\text{quark}(p^2)/\Pi_\text{ghost}(p^2)$ as a function of the external momentum $p^2$.

For $N_f=0$ the quark loop decouples, and we are left with the gluon and ghost loop. As we increase $N_f$ the quark loop enters the system with the same sign as the gluon loop. The quark loop increases in strength and the gluon loop decreases for increasing $N_f$. The sum of both matches the magnitude of the ghost loop for all $N_f<$\Nfcrit, and again there is the pattern of large cancellation. Already at $N_f \lessapprox 0.8$ the gluon loop changes sign not only in the deep IR but in the whole non-perturbative  regime. We observe this pattern for also for the other gauge groups studied in sec.\ \ref{sec:other_groups} and report the number of fermions for which the gluon loop is approximately vanishing in the IR in table \ref{tab:vanishing_gluon_loop}.\footnote{
We stress that any quoted uncertainties  in this and the following tables only refer to the numerical uncertainties within the specific truncation scheme and for the specified vertex models. Specifically, it excludes any estimation of the systematic uncertainties associated with our truncation and/or models.} 
As we approach the conformal window the cancellations of quark and gluon loop on the one hand and the ghost loop on the other hand become less pronounced, and eventually the gluon and quark loops dominate after the phase transition. We conclude that at no point in a physical system where $N_f$ assumes integer values the quark loop can be assumed to have only small contributions to this system.

\begin{table}
    \centering
    \begin{tabular}{|l|l|}
      \hline \hline 
       & $N_f^{0}$  \\ 
      \hline \hline 
      \SU{2} & 0.50(10)  \\
      \SU{3} & 0.75(5)   \\
      \SU{4} & 1.00(10)  \\
      \SU{5} & 1.25(10)   \\
      \hline \hline
    \end{tabular}

 \smallskip
 
    \begin{tabular}{|l|l|}
      \hline \hline 
      \Sp{2} & 0.50(10) \\
      \Sp{4} & 0.75(5)  \\
      \Sp{6} & 1.00(10) \\
      \Sp{8} & 1.2(2)   \\
      \hline \hline
    \end{tabular}

\smallskip

    \begin{tabular}{|l|l|}
      \hline \hline 
      \SO{6} & 0.50(10) \\
      \SO{7} & 0.60(10) \\
      \SO{8} & 0.75(10) \\
      \SO{9} & 0.90(10) \\
      \hline \hline
    \end{tabular}
    \caption{Number of flavors $N_f^0$ for which the gluon loop is vanishing in the infrared and mid-momentum regime.}
    \label{tab:vanishing_gluon_loop}
\end{table}

\subsection{Other gauge groups} 
\label{sec:other_groups}
So far we have focused on the case of \SU{3}\ gauge theory with fermions in the fundamental representations. However, the structure of the DSEs allows us to study other gauge groups and in principle also other fermion representations. This allows us to identify common features between different strongly interacting, confining theories. As stated in \ref{sec:results_vertex}, we do not believe that our results for \Nfcrit\ and the IR exponent $\rho$ provide a quantitatively correct description of extending the physics of QCD to the conformal window. However, by comparing different theories we might be able to make much more robust statements about the relative size of \Nfcrit and $\rho$. These results provide then novel non-perturbative statements that complement others calculations \cite{Lee:2020ihn} and can be of interest for BSM model building.
\begin{table}
    \centering
    \begin{tabular}{|l|l|l|c|}
      \hline \hline 
             & $N_f^{AF}$ & \Nfcrit & $\rho$ \\ 
      \hline \hline 
      \SU{2} & 11   & 2.70(5) & 0.222(3) \\
      \SU{3} & 16.5 & 4.76(2) & 0.174(4) \\
      \SU{4} & 22   & 6.58(5) & 0.163(3) \\
      \SU{5} & 27.5 & 8.35(5) & 0.161(4) \\
      \hline \hline
      \Sp{2} & 11   & 2.70(5) & 0.224(1) \\
      \Sp{4} & 16.5 & 4.50(5) & 0.192(5) \\
      \Sp{6} & 22   & 6.15(5) & 0.181(6) \\
      \Sp{8} & 27.5 & 7.95(5) & 0.173(5) \\
      \hline \hline
      \SO{6} & 11    & 3.95(5)  & 0.121(3) \\
      \SO{7} & 13.75 & 4.80(10) & 0.126(2) \\
      \SO{8} & 16.5  & 5.65(4)  & 0.131(2) \\
      \SO{9} & 19.25 & 6.40(5)  & 0.135(2) \\
      \hline \hline 
    \end{tabular}
    \caption{The number of critical Dirac fermions in the fundamental representation for both \SU{N}, \Sp{2N}\ and \SO{N}\ gauge theory with a bare ghost-gluon vertex. Additionally, we report the infrared exponent $\rho$ according to Eq.~\eqref{eq:IR_exponent}. The uncertainties refer only to the fixed truncation and fixed vertex models. Systematic errors due to the truncation and model-dependence are \textit{not} taken into account.}
    \label{tab:results}
\end{table}
\begin{table}
    \centering
    \begin{tabular}{|l|c|c|c|}
      \hline \hline 
             & $N_f^{AF}$ & \Nfcrit & $\rho$ \\ 
      \hline \hline 
      \SU{2} & 11    & 2.65(5) & 0.225(2) \\
      \SU{3} & 16.5  & 4.76(2) & 0.178(4) \\
      \SU{4} & 22    & 6.58(3) & 0.166(3) \\
      \SU{5} & 27.5  & 8.35(5) & 0.163(4) \\
      \hline \hline
      \Sp{2} & 11    & 2.65(5) & 0.225(2) \\
      \Sp{4} & 16.5  & 4.50(5) & 0.196(4) \\
      \Sp{6} & 22    & 6.20(5) & 0.182(5) \\
      \Sp{8} & 27.5  & 7.95(5) & 0.176(5) \\
      \hline \hline
      \SO{6} & 11    & 3.90(10)& 0.121(7) \\
      \SO{7} & 13.75 & 4.80(10)& 0.127(2) \\
      \SO{8} & 16.5  & 5.68(3) & 0.132(2) \\
      \SO{9} & 19.25 & 6.35(5) & 0.138(2) \\
      \hline \hline 
    \end{tabular}
    \caption{The same data as in table \ref{tab:results} but with a dynamical, self-consistent ghost-gluon vertex. Within the employed
    truncation we observe no change in \Nfcrit\ from the inclusion of the ghost-gluon vertex.}
    \label{tab:resultsGGV}
\end{table}

In table \ref{tab:results} the corresponding results for the critical number of Dirac fermions \Nfcrit~ and the IR exponent $\rho$ are given for $\SU{N=2,3,4,5}$; as well as for $\Sp{2N=2,4,6,8}$ and $\SO{N=6,7,8,9}$. This is motivated by the fact that within the \SU{N} and \Sp{2N} series the  value of $N_f^{AF}=11C_A/(4T)$ is mutually equal. Additionally, we have chosen to start with \SO{N=6} gauge group because its value of $N_f^{AF}$ matches those of \SU{2} gauge theory.
We find that in any of the considered theories, the ghost-gluon-vertex leads to no significant change in the value of \Nfcrit. Neither did it affect the value of the exponent governing the IR power laws $\rho$. As in the case of $\SU{3}$, the vertex shows only small deviations from tree-level as the number of fermions is increased. 

For the symplectic gauge theories we observe a slightly smaller \Nfcrit \ than for the corresponding \SU{N} gauge theory with equal $N_f^{AF}$, whereas the values for corresponding \SO{N} are noticeably larger.
For the exponent $\rho$ our numerical results suggest that the respective values are decreasing with $N$ for the \SU{N}\ and \Sp{2N}\ series. They appear to be decreasing slightly slower for the \Sp{2N} series. 
In contrast, the exponent $\rho$ increases with $N$ for the orthogonal groups \SO{N}, however, the values of $\rho$ stay still significantly below the analogous ones for the \SU{N}\ and \Sp{2N}\ series.

\begin{figure}
    \centering
    \includegraphics[width=.47\textwidth]{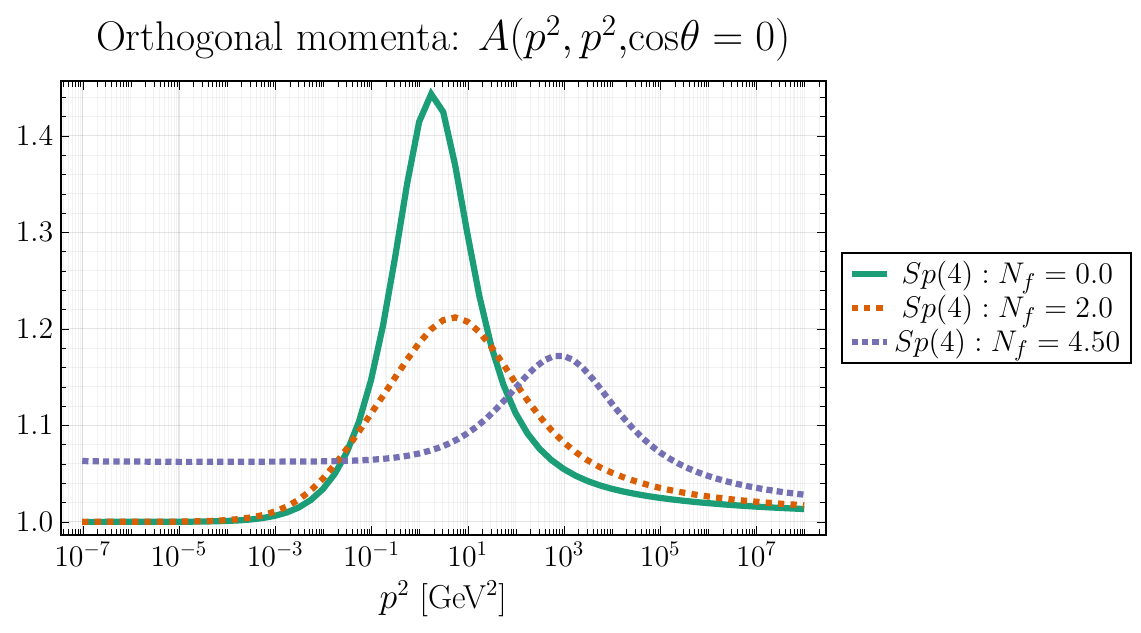}
    \includegraphics[width=.47\textwidth]{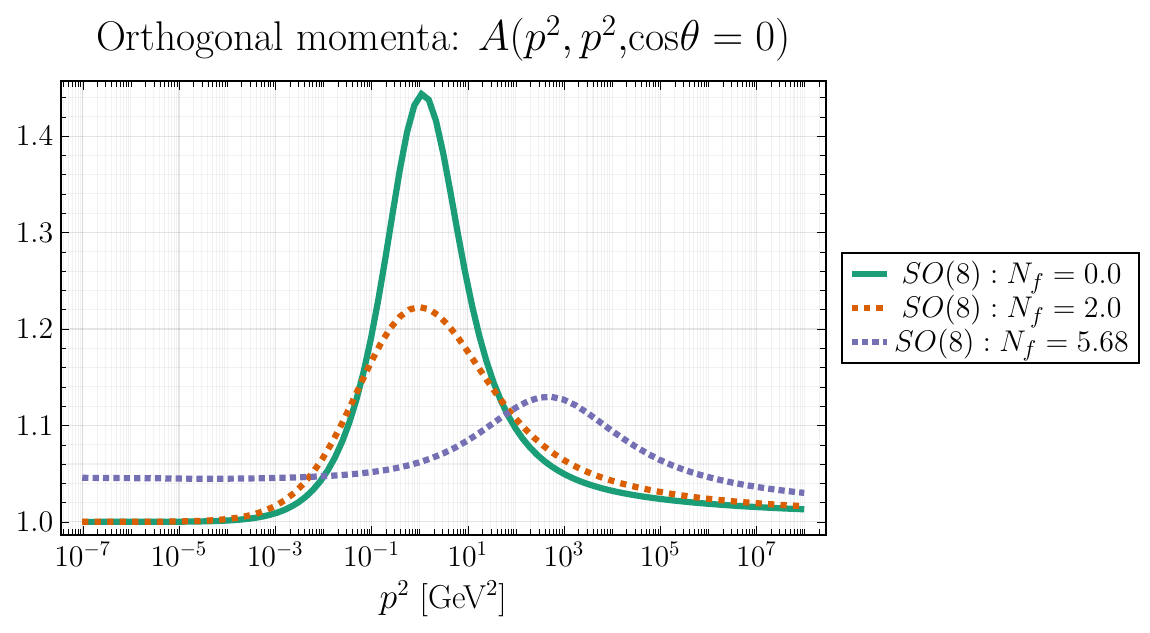}
    \caption{Ghost-gluon vertex for orthogonal momenta and for the gauge groups \Sp{4} (upper panel) and \SO{8} (lower panel) for different values of $N_f$.}
    \label{fig:GGV_Sp4SO8}
\end{figure}
\begin{figure}
    \centering
    \includegraphics[width=.47\textwidth]{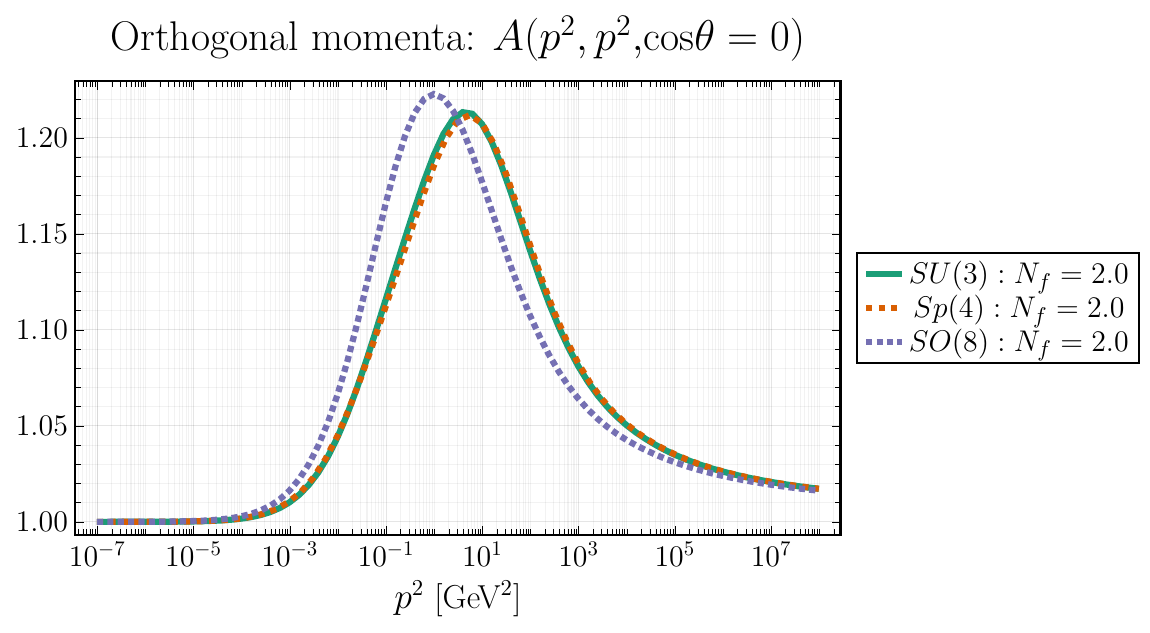}
    \caption{Ghost-gluon vertex for $N_f= 2$ for the gauge groups \SU{3}, \Sp{4} and \SO{8} for orthogonal momenta. }
    \label{fig:GGV_Nf2}
\end{figure}
In Fig.\ \ref{fig:GGV_Sp4SO8} we display the form factor of the ghost-gluon vertex for the gauge groups \Sp{4} and \SO{8} for orthogonal momenta 
and different values of $N_f$. We observe here the very same pattern as for the \SU{3} gauge group. How similar the ghost-gluon vertex form factors for these three gauge groups
are becomes evident from Fig.\ \ref{fig:GGV_Nf2}. Note that for these three groups $N_f^{AF}$ is identical, and the values for \Nfcrit in the employed truncation deviate only mildly. 
Choosing other combinations with identical $N_f^{AF}$ as, {\it e.g.}, \SU{2}=\Sp{2} and \SO{6} or the triple \SU{4}, \Sp{6} and \SO{10}, provides completely analogous comparisons.

\begin{figure}
    \centering
    \includegraphics[width=.47\textwidth]{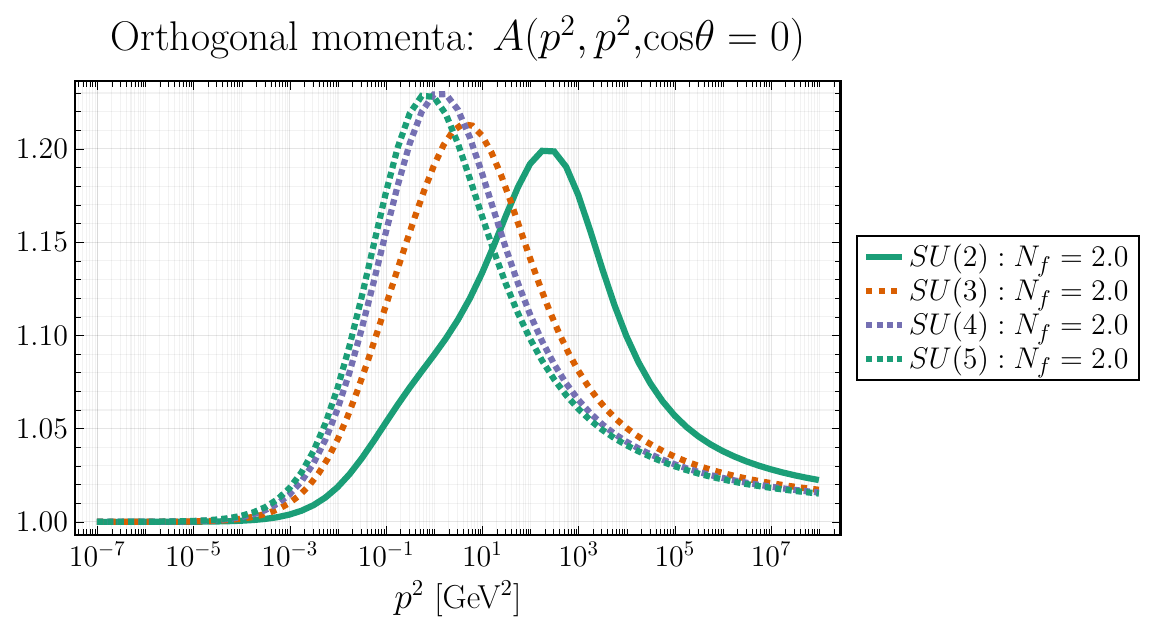}
    \includegraphics[width=.47\textwidth]{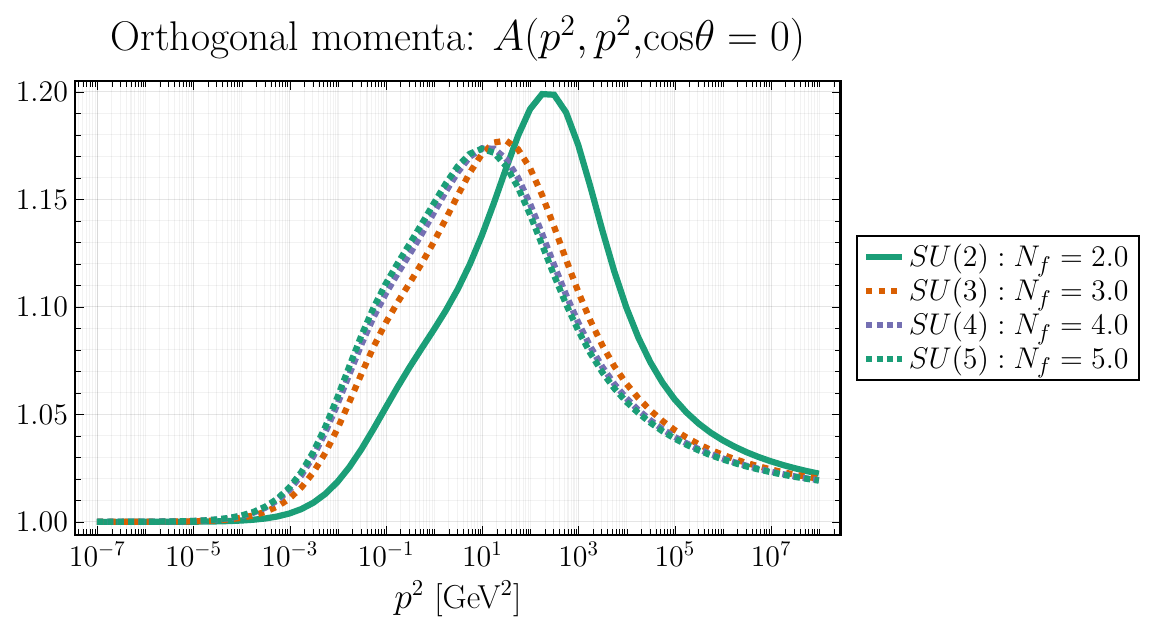}
    \caption{Ghost-gluon vertex for orthogonal momenta and for the gauge groups \SU{N}, $N=2,3,4,5$ and $N_f=2$(upper panel) and \SU{N}, $N=2,3,4,5$ and $N_f=N$. }
    \label{fig:GGV_largeN}
\end{figure}
In Fig.\ \ref{fig:GGV_largeN} the form factor of the ghost-gluon vertex is plotted for orthogonal momenta and the gauge groups \SU{N} for $N = 2 \ldots 5$. Hereby in the upper panel
$N_f=2$ is kept fixed. One can infer that as expected the unquenching effects become smaller with increasing $N$. One additionally sees that in the 't Hooft limit one obtains
one unique function for the form factor of the ghost-gluon vertex and that this function is reached very quickly such that the \SU{4} ghost-gluon vertex represents the limit already very precisely.
In the lower panel the ghost-gluon vertex for same gauge groups but with $N_f=N$ are displayed. For this case the unquenching effects are of a similar size, and more interestingly the Veneziano 
limit ($N\to \infty$ with fixed $N_f/N$) is assumed even faster than the 't Hooft limit.

\section{Conclusion}
\label{sec:Conclusion}

In summary, we have studied the ghost-gluon vertex as well as the influence of this vertex on the gluon, ghost and quark propagators with $N_f$ massless Dirac fermions in the fundamental representation. For all studied values of $N_f$ the dressing of the ghost-gluon vertex  remains small, and  with increasing $N_f$ it becomes even smaller than in the quenched case. Moreover, the infrared behavior of this vertex provides further evidence for the existence of an infrared fixed point above \Nfcrit. 

Additionally, we have investigated the relative size of the gluon and the quark loop in the gluon DSE with the result that the quark loop becomes of roughly equal magnitude as the gluon and the ghost loops even for quite small values of $N_f$. For $N_f\lessapprox$ \Nfcrit \ the sum of the quark and the gluon loop on the one hand and the ghost loop on the other hand cancel quite precisely in the infrared and the (non-perturbative) mid-momentum regime.

A comparison between different gauge groups shows some systematic differences between \SU{N}, \Sp{2N}\ and \SO{N}\ gauge theories. However, taking into account the obtained evidence that the 
ghost-gluon vertex converges uniformly in the 't Hooft as well as in the Veneziano limit with the asymptotic functions being numerically accurate already at quite small values of $N$ we can safely
conclude that in the Landau gauge the deviation of the ghost-gluon vertex from its tree-level values is small for all $N$ and all $N_f<N_F^{AF}$ ({\it i.e.}, for all $N_f$ such that asymptotic freedom 
is respected). 

Consequently, the self-consistent inclusion of the ghost-gluon vertex has no qualitative and only little quantitative impact on the gluon, ghost and quark propagators for all asymptotically free 
 \SU{N}, \Sp{2N}\ and \SO{N}\ gauge theories with an otherwise unconstrained number of fermions in the fundamental representation.

\bigskip 

\section*{Acknowledgments}
We thank Madeleine Adrien, Eduardo Ferreira, Markus Q. Huber, Axel Maas, Joachim Pomper, H\`elios Sanchis-Alepuz and Georg Wieland for helpful discussions. FZ has been supported in part by the Austrian Science Fund research teams grant STRONG-DM (FG1) and the STFC Grant No. ST/X000648/1.

\bigskip

{\bf Open Access Statement}---For the purpose of open access, the authors have applied a Creative Commons Attribution (CC BY) licence  to any Author Accepted Manuscript version arising.

{\bf Research Data Access Statement}---The data generated for this manuscript can be downloaded from Ref.~\cite{data_release}, and the software used to generate it is similarly available from Ref.~\cite{code_release}.

\bigskip

\bibliographystyle{apsrev4-1}
\bibliography{refs}

\end{document}